# Optical Activity of a Dilute Dielectric Medium: Pasteur and the Molecular Symmetries

M. Cattani[(*)] and J. M. F. Bassalo[(**)]

[(*)]Instituto de Física da Universidade de São Paulo. C.P. 66318, CEP 05315-970
São Paulo, SP, Brasil.

E-mail < mcattani@if.usp.br >

[(**)]Fundação Minerva, R.Serzedelo Correa 347,1601 - CEP 66035-400 Belém-Pará Brasil.

E-mail <bassalo@amazon.com.br>

Abstract
    In 1848 Pasteur conjectured that the rotation of the polarization plane of the light in a dilute dielectric medium is generated by the molecular symmetries of the medium where the light propagates. Our objective is to show that Pasteur hypothesis was correct using basic knowledge of electromagnetism and quantum mechanics. In Sections 2-5 we present a brief review of the fundamental concepts of the electromagnetism necessary to study the optical activity. In Sections 6-8 using the quantum mechanical perturbation theory and taking into account the molecular symmetries we calculate the optical activity of the medium. It will be shown that the theoretical predictions that are in good agreement with the experimental results give support to the Pasteur hypothesis.

**Keywords:** *optical activity, molecular symmetries.*

**(1)Introduction: Molecular Optical Activity and Pasteur.**
    In a preceding paper [1], where was analyzed the effect of the weak interactions in the chiral biochemistry, we made a brief retrospect of the first experiments on the optical activity of tartaric acid crystals. These crystals appear in the fermentation process of grapes in the production of wine. These crystals named sodium ammonium tartrate (SAT) that are represented by the formula $NaOOC-CHOH-CHOH-COONH_4$ have the basic structure composed by the tartaric acid ($C_4H_6O_6$) also called racemic acid (from the Latin word "racemus" which means grape). A detailed historical description of the studies about the optical activity of the tartaric and other crystals is found in the excellent book "Optical Rotatoty Power" of T M. Lowry [2]. We suggest also the lecture of the paper "Optical Activity and Molecular Dissymmetry" de S.F.Mason [3]. In these references the reader will find the description of the most significant works about the optical activity since the pioneer works of, for instance, Christiaan Huygens (1690), Jean Baptiste Biot (1812) and Augustin Fresnel (1825)

Pasteur, between 1848 and 1850, with 26 years old, analyzing in details samples of crystals of SAT obtained in fermentation tanks of wines [2] he verified that they were composed by two different types of small crystals that were mirror images one of the other, as occurs with the right and left hands (see Fig.1). Objects that are not identical with their mirror images have *chiral* (from the Greek word "keir" which means hand) symmetry. In the literature the *left-handed* or simply *left* is indicated by symbol L (levogire) and the *right-handed* or simply *right* is indicated by R or D (destrogire).

Laboriously, with help of one magnifying glass and tweezers he separated the crystals *right* (R or D) and *left* (L*)* in two small portions. After he verified the separate aqueous solutions of each one of these different portions presented opposed optical activities: the solution of one of the portions turned the polarization plane of the light in the *left* sense (or clockwise sense) and the other in the *right* sense (or counter-clockwise sense). He also showed that solutions obtained with equal number of L and R crystals do not altered the polarization plane. The solution composed with 50% of L crystals and 50% of R crystals is called *racemic* solution. Pasteur in 1848 [2] created a neologism to baptize the L and R crystals calling them *dissymmetrics.* In Fig.1 is shown the two types of crystals of TSA, one L– rotatory and the other R – rotatory.

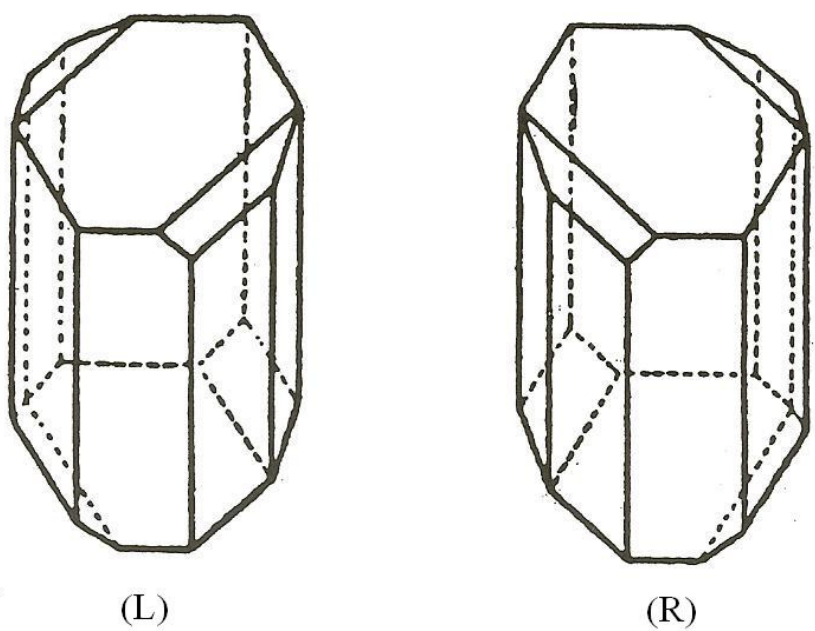

(L)　　　　　　　　　　(R)

Figura 1

Figura 1. Dissymmetric crystals L and R of sodium ammonium tartrate (SAT).

As the differences in the light polarization rotations occurred in aqueous solutions Pasteur conjectured that the rotations were not produced by the crystals but instead generated by the tartrate molecules since the crystals were dissolved in the water. That is, it ought to have a molecular *dissymmetry* that would be responsible by the different rotations of the light polarization. He classified these molecules in two types: *left-hand* (levogire) and *right-hand* (destrogire). Today these molecules, known as *chiral* molecules, are called *enantiomers* and are of two types: L-enantiomer and R- enantiomer. These molecules have the property that there is no superposition between their structural representations and their respective mirror images, as occurs with the left and right hands. Due to this discovery he was honored with the "Légion d'Honneur Francaise."

In Fig.2 are shown the dissymmetric structures L and R of the tartaric acid that are the elementary structures of the SAT crystals.

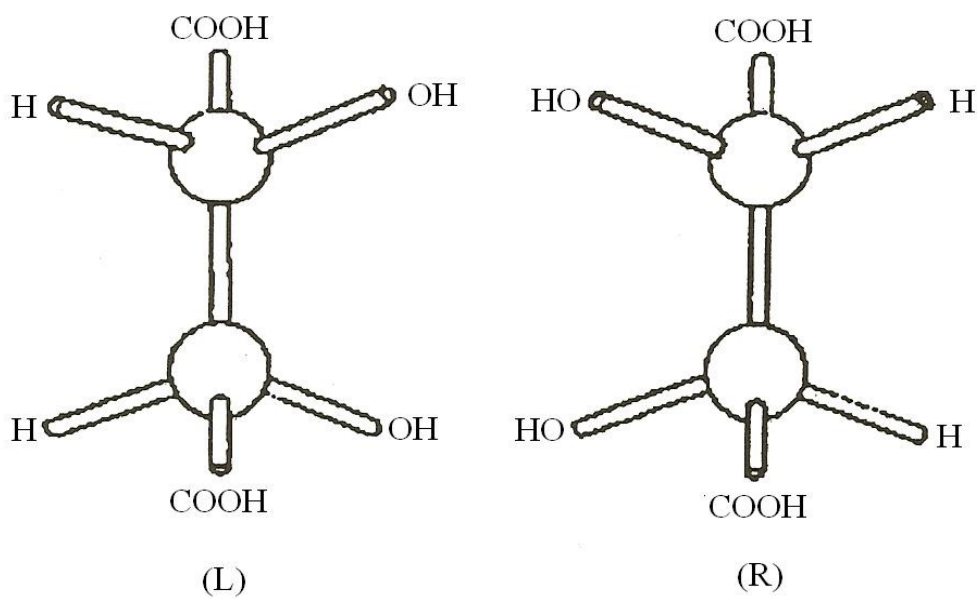

Figura 2

Fig.2. Dissymmetric Structural Forms L and R of the tartaric acid ($C_4H_6O_6$).

Pasteur, later, after analyzing several other experiments [1] where were found dissymetries he understood that there was something more profound in the "left-right" phenomena. With this conviction he presented in the French Academy of Sciences the famous conjecture: "The Universe is dissymmetric". Only about one century later, in 1957, it was verified that Pasteur was correct with the experimental verification of the parity non-conservation in the weak interactions [1,4].

It is important to note that, in 1815, Biot [2,3] that was professor of Pasteur discovered that many substances in nature (like, for instance, sugar and camphor) have optical activity not only crystallized but also in solutions. From this he also inferred in 1817 that the asymmetry that produces the rotation of the polarization plane of the light ought to be inside the molecules of the substances. Fresnel [2,3] to explain the two kinds of polarization, LCP (left circular polarization) and RCP (right circular polarization) of an optically active medium postulated in 1825 that the molecules of the medium ought to have helicoidal forms, ones with left screw and others with left screw. Today these molecular characteristics are well known and studied from the point of view of their symmetries and physical and chemical properties. These studies belong to a branch of chemistry known as Optical Isomery [3].

There are two famous enantiomers that have two optical isomers: the thalidomide and the limonene. One enantiomer of the thalidomide cures morning sickness, the other causes birth defects. The limonene is a compound found in lemons and oranges. One enantiomer has a lemon flavor, the other an orange flavor.

In Fig.3 are shown the enantiomorphs hexagonal crystals of quartz commonly used to study the optical activity [2].

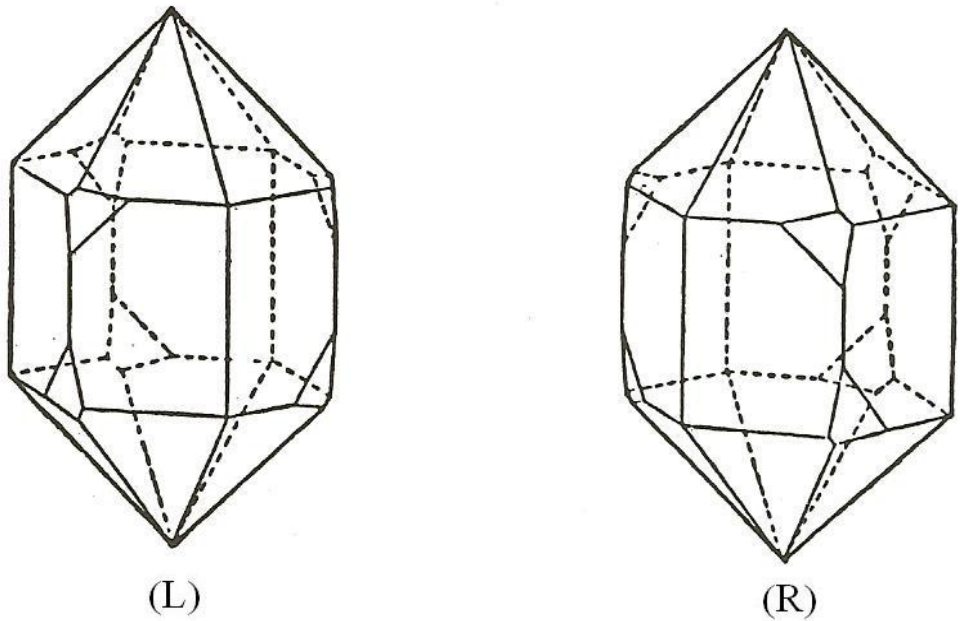

Figura 3

Fig.3. The two enantiomorphs or dissymmetrics forms, L and R, of the hexagonal quartz crystal.

In Section 2 are shown the Maxwell equations in the CGS system that will be adopted in this paper. In Sections 3 and 4 we will assume *ab initio* that the refraction index $n_l$ and $n_r$ of the left circularly polarized (LCP) and right circularly polarized RCP waves, respectively, are complex. That is, $n_l = n_l + i\,\chi_l$ and

$n_r = n_r + i\,\chi_r$, where n and χ are real. With these complex functions we will show that the *rotatory power* [φ] is given by [φ] = (π/λ)($n_l - n_r$) where λ is light wavelength in the vacuum and that the circular dichroism or Cotton Effect is measured by the *ellipticity* $\Psi$ given by $\Psi = (\pi/\lambda)(\chi_l - \chi_r)$. In Section 5 using the electromagnetic theory and the classic atomic model we calculate the refraction index $n_l$ and $n_r$ of the LCP and LCP waves when the dielectric medium is submitted to a constant external magnetic field **B** (*Faraday Effect* or *Magnetic Birefringence*). In Section 6 we will generalize the Maxwell equations assuming that the medium is composed by molecules with internal structures which symmetries will be responsible by the rotation of the polarization plane and by the absorption of the intensity of the electromagnetic field. That is, there we will see how the microscopic molecular symmetries will affect the macroscopic fields **E** and **B**. There we will calculate [φ] and $\Psi$ using the quantum mechanics and taking into account the symmetries of the molecules of the medium. We verify that there is a good agreement between theory and experiment proving, in this way, the Pasteur hypothesis.

**(2) Maxwell Equations in Dielectric Medium.**
Before to study the rotation of the plane polarization of the light in a dielectric medium let us make a brief review of the Maxwell equations [5-7] and their main properties in this medium. In the CGS unity system the Maxwell equations, in the general case, are given by:

$$\text{div } \mathbf{D} = \rho_{real} \text{ [ Gauss]}, \quad \text{rot } \mathbf{E} = -(1/c)(\partial \mathbf{B}/\partial t) \quad \text{[Faraday]}$$
$$\text{div } \mathbf{B} = 0, \quad \text{rot } \mathbf{H} = \mathbf{j}_{real} + (1/c)(\partial \mathbf{D}/\partial t), \quad \text{[Ampère]}$$
(2.1)

that in absence of real charges ($\rho_{real} = 0$) and real currents ($\mathbf{j}_{real} = 0$) become

$$\text{div } \mathbf{D} = 0, \quad \text{rot } \mathbf{E} = -(1/c)(\partial \mathbf{B}/\partial t),$$
$$\text{div } \mathbf{B} = 0, \quad \text{rot } \mathbf{H} = (1/c)(\partial \mathbf{D}/\partial t).$$
(2.2)

The vectors **D** and **B** that are defined by $\mathbf{D} = \mathbf{E} + 4\pi\mathbf{P}$ and $\mathbf{B} = \mathbf{H} + 4\pi\mathbf{M}$, where **P** is the electric polarization vector and **M** is the magnetic polarization vector. For an isotropic medium **P** and **M** are given, respectively, by $\mathbf{P} = \kappa \mathbf{E}$ and $\mathbf{M} = \kappa' \mathbf{E}$ where κ (electric susceptibility) and κ´(magnetic susceptibility) are scalars. In this way, $\mathbf{D} = (1 + 4\pi \kappa) \mathbf{E} = \varepsilon \mathbf{E}$ and $\mathbf{B} = (1 + 4\pi \kappa') \mathbf{H} = \mu \mathbf{H}$, where ε e μ are, the dielectric constant and the magnetic permeability, respectively, of the medium. The refraction index of the medium is defined by $n^2 = \varepsilon\mu$ and the velocity V of propagation of a wave is $V = c/n = c/\sqrt{\varepsilon\mu}$. For a weakly magnetic medium we have k´ << 1. In this case we can put $\mu \approx 1$, $n^2 \approx \varepsilon$, $V = c/\sqrt{\varepsilon}$ and $\mathbf{B} = \mathbf{H}$.

The polarization $\mathbf{P} = \kappa \mathbf{E}$ where **E** is the field created by all charges, real and induced, and $\mathbf{D} = \varepsilon \mathbf{E}$ is created by real charges. For a dense medium, due to the charge polarization effect, the effective field $\mathbf{E}_{ef}$ that acts on a molecule is given by $\mathbf{E}_{ef} = \mathbf{E} + (4\pi/3)\mathbf{P}$. As the molecular polarizability α is defined by $\mathbf{P} = N \alpha \mathbf{E}_{ef}$ we get

$$\mathbf{P} = \kappa \mathbf{E} = N \alpha \mathbf{E}_{ef} = N \alpha [ \mathbf{E} + (4\pi/3)\mathbf{P} ] = N \alpha (1 + 4\pi\kappa/3) \mathbf{E} \quad (2.3),$$

from which we obtain,

$$\kappa = N\alpha/(1 - 4\pi N\alpha/3) \tag{2.4}$$

Remembering that $\varepsilon = 1 + 4\pi\kappa$, that is, $\kappa = (\varepsilon - 1)/4\pi$ we get, using (2.4),

$$\alpha = (3/4\pi N)(\varepsilon - 1)/(\varepsilon + 2) \tag{2.5},$$

which is the Clausius-Mossotti equation.

When the magnetization is negligible, $n^2 = \varepsilon\mu \approx \varepsilon$ and (2.5) becomes

$$(n^2 - 1)/(n^2 + 2) \approx (4\pi N/3)\alpha \tag{2.6}.$$

In a dilute medium, that is, when $4\pi N\alpha/3 \ll 1$, using (2.4) and the relation $\varepsilon = 1 + 4\pi\kappa$ we have

$$\kappa \approx N\alpha \quad \text{and} \quad \varepsilon \approx 1 + 4\pi N\alpha, \tag{2.7}$$

In this way, for a dilute medium with a very weak magnetization we see that

$$\varepsilon \approx n^2 \approx 1 + 4\pi N\alpha. \tag{2.8}.$$

For an homogeneous medium, where k, k´, ε and μ are constants, the Eqs.(2.2) can also be written as:

$$\text{div } \mathbf{E} = 0, \quad \text{rot } \mathbf{E} = -(1/c)(\partial \mathbf{B}/\partial t),$$
$$\text{div } \mathbf{B} = 0, \quad \text{rot } \mathbf{B} = (\varepsilon\mu/c)(\partial \mathbf{E}/\partial t). \tag{2.9}$$

In these conditions we have the following wave equations

$$\Delta \boldsymbol{\Phi} - (1/V^2)(\partial^2 \boldsymbol{\Phi}/\partial t^2) = 0, \tag{2.10}$$

where $\boldsymbol{\Phi} = \mathbf{E}$ or $\mathbf{B}$ and $V = c/n = c/\sqrt{\varepsilon\mu}$.

**(3) Rotation of the Polarization Plane in a Dielectric Medium.**

We study here the light propagation assuming that the medium is isotropic, homogeneous and with negligible magnetization. In these conditions the fields $\mathbf{E}(\mathbf{r},t)$ and $\mathbf{B}(\mathbf{r},t)$ propagate in the medium obeying the wave equations (2.9)-(2.10) with $n^2 \approx \varepsilon$, that is, $n \approx \sqrt{\varepsilon}$, and $V \approx c/\sqrt{\varepsilon}$. The analysis done in this Section is performed following the Condon's paper [12]. It can also be seen in books of optics [13-16] where there are many figures showing clearly the superposition of the waves, rotation of the polarization planes, circular and elliptic polarizations.

Let us assume that $\mathbf{E}(\mathbf{r},t)$ and $\mathbf{B}(\mathbf{r},t)$ propagate along an axis z perpendicular to the plane (x,y) of the page, pointing in the reader sense. We define by ***x, y*** and ***z*** the unitary vectors along the axis x, y and z, respectively. The unitary vectors (***x, y, z***) form a basis of a "right-handed" system of coordinates. The fields **E** and **B** that are in the (x,y) plane are given by:

$$\mathbf{E} = \mathrm{Re}\{\mathbf{E}_o \exp(i\Phi)\} \qquad \text{and} \qquad \mathbf{B} = \mathrm{Re}\{\mathbf{B}_o \exp(i\Phi)\} \quad (3.1),$$

where the wave phase $\Phi = \omega(t - n\mathbf{z}\cdot\mathbf{r}/c) = \omega t - \mathbf{k}\cdot\mathbf{r}$ and Re{...} means that we are taking the real part of the function that is inside the bracket, $\mathbf{E}_o$ and $\mathbf{B}_o$ are constant vectors, $\omega = 2\pi f$ is the angular frequency and f the frequency measured in hertz of the wave (**E**, **B**). The wave vector **k** is given by $\mathbf{k} = (2\pi/\lambda)\mathbf{z}$, $\lambda = 2\pi c/\omega$ is the wavelength of the wave in the vacuum and $\lambda_n = \lambda/n$ is the wavelength in the dielectric medium.

(3.a) RCP wave.

Let us suppose that the amplitude $\mathbf{E}_o$ is given by $\mathbf{E}_r = e_o(\mathbf{\textit{x}} + i\mathbf{\textit{y}})$, $e_o$ being real. In this case the real part of **E** is given by

$$e_o[\mathbf{\textit{x}}\cos(\Phi) - \mathbf{\textit{y}}\sin(\Phi)]. \qquad (3.2).$$

When $\Phi = 0$ (assuming that at the initial instant t = 0 the initial position is $\mathbf{r} = 0$) the vector **E** is parallel to ***x*** and as $\Phi$ increases **E**, seen by the reader, turns in the counter-clockwise sense. This wave will be indicated by RCP or CP+ (wave with positive ellipticity).

**(**3.b) LCP wave.

In this case the wave amplitude $\mathbf{E}_o$ is given by $\mathbf{E}_l = e_o(\mathbf{\textit{x}} - i\mathbf{\textit{y}})$ and its real part is given by,

$$\mathbf{E}_l = e_o[\mathbf{\textit{x}}\cos(\Phi) + \mathbf{\textit{y}}\sin(\Phi)] \qquad (3.3).$$

When $\Phi = 0$ (taking $\mathbf{r} = 0$ at t = 0) the vector **E** is parallel to ***x*** and as $\Phi$ increases **E**, seen by the reader, turns in clockwise-sense. This wave will be indicated by LCP or CP− (wave with negative ellipticity).

(3,c) LP wave created by LCP and RCP waves with equal amplitudes.

Let us consider the superposition of two circularly polarized waves with equal amplitudes**.** One of them RCP, with phase θ, given by
$\mathbf{E}_R = \mathrm{Re}\{\mathbf{E}_r \exp(i\theta)\exp(i\Phi)\}$ and other LCP with phase −θ, given by
$\mathbf{E}_L = \mathrm{Re}\{\mathbf{E}_l \exp(-i\theta)\exp(i\Phi)\}$. In this case the resultant wave $\mathbf{E} = \mathbf{E}_R + \mathbf{E}_L$ is given by

$$\mathbf{E} = e_o\,\mathrm{Re}\{(\mathbf{\textit{x}} + i\mathbf{\textit{y}})\exp(i\theta) + (\mathbf{\textit{x}} - i\mathbf{\textit{y}})\exp(-i\theta)\} = 2e_o\{\mathbf{\textit{x}}\cos(\theta) - \mathbf{\textit{y}}\sin(\theta)\} \quad (3.4).$$

The above equation (3.4) shows that the superposition of the RCP and LCP waves generates a linearly polarized wave (LP). When θ = 0 the LP wave has a polarization plane parallel to **x**. When θ > 0 the polarization plane is rotated by an angle θ relatively to **x** in the clockwise-sense.

(3.d) Propagation RCP and LCP waves with different refraction indexes.

Now let us assume that the RCP and LCP waves have different propagation velocities in the medium, that is, $n_r \neq n_l$. This difference in the refraction indexes will be explained in Section 6 taking into account the molecular symmetries of the medium. In addition we assume that the RCP and LCP waves have the same amplitude and that at t = 0 we have **z·r** = 0. After a time t the waves travel a distance L so that **z·r** = L with constant amplitudes. In this time interval t the phases Φ of the two components will be given by:

$$\Phi_r = \omega(t - n_r L/c) \quad \text{and} \quad \Phi_l = \omega(t - n_l L/c) \quad (3.5).$$

In this way the resultant field **E** at the time t and **z·r** = L becomes

**E**(**r**,t) = Re{**E**$_R$ exp(iΦ$_r$) + **E**$_L$ exp(iΦ$_l$)} =

= $e_o$ Re{ ( **x** + i **y** ) exp(iθ) exp(iΦ$_r$) + (**x** – i **y**) exp(–iθ) exp(iΦ$_l$) }.

Defining Φ = ω[t – ($n_l$ + $n_r$)L/2c] and δ = π ($n_l$ – $n_r$)L/λ, where λ = cf = 2πc/ω is the wavelength in vacuum, the equation seen above becomes:

**E**(**r**,t) = $e_o$ Re{ ( **x** + i **y** ) exp(iθ) exp(i δ) + (**x** – i **y**) exp(-iθ) exp(–iδ) } exp(iΦ).

Assuming now that the light entered LP parallel to **i**, that is, that θ = 0 we get:

$$\mathbf{E}(\mathbf{r},t) = e_o \text{Re}\{ (\mathbf{x} + i\mathbf{y}) \exp(i\delta) + (\mathbf{x} - i\mathbf{y}) \exp(-i\delta)\} \exp(i\Phi) \quad (3.6).$$

In this way we see that the amplitude of **E**(**r**,t) given by (3.6) is equal to

$$e_o \text{Re}\{(\mathbf{x} + i\mathbf{y}) \exp(i\delta) + (\mathbf{x} - i\mathbf{y}) \exp(-i\delta)\} = 2 e_o \{\mathbf{x} \cos(\delta) - \mathbf{y} \sin(\delta)\} \quad (3.7).$$

This result shows that the polarization plane that entered LP in the plane parallel to **x** was turned by an angle δ relatively to **x**. The rotation δ is in the clockwise or in the counter-clockwise sense if δ > 0 or δ < 0, respectively. Since δ = π ($n_l$ – $n_r$)L/λ the δ signal will depend of the difference ($n_l$ – $n_r$).

Thus, a wave that entered LP in the medium, composed by two CP waves with equal amplitudes but with different refraction indexes, $n_l$ e $n_r$, after to cover a distance L, continues LP but its polarization plane is rotated by an angle δ. This rotation angle per unit of distance of the trajectory is the *rotatory power* [α] = δ/L given by:

$$[\alpha] = \delta/L = (\pi/\lambda)(n_l - n_r). \quad (3.8)$$

The parameter [α] is measured in radian/m or degree/m. Since λ is very small compared with the distances L the quantity L/λ is very large. Consequently, despite the difference $n_l - n_r$ be very small compared with the unity, around one part in one million, considerable values of the rotation δ are produced. Note that the sense of the rotation δ ( δ > 0 is clockwise sense and δ < 0 is counter-clockwise) is given by the CP light that more quickly propagates in the medium since V = c/n. Fresnel [3] was the first one to obtain, in 1825, the Eq.(3.8).

**(4) Attenuation of LCP and RCP Waves. Cotton Effect. CircularDichroism.**

As well known [2,9-12] when there is a rotation δ of the polarization plane according to (3.8) there is also an absorption of the intensities of the LCP and RCP waves along the trajectory of the light. The attenuation of the intensities of the LCP and RCP waves are different. This absorption process that was discovered by Cotton in 1896 is called Cotton Effect or Circular Dichroism [2,3,8-16]. As will be shown in Section 8 the rotatory power [α] and the circular dichroism are intrinsically connected. To explain the dichroism we need to consider the complex refraction index *n*. Thus, we will follow the calculations performed in Section 3 substituting the real refraction indexes $n_l$ and $n_r$, of the RCP and LCP waves, by the complex refraction indexes, $n_l = n_l + i\chi_r$ and $n_r = n_r + i\chi_l$, respectively. In these conditions due to the absorption coefficient χ the amplitudes of the RCP and LCP waves will decrease, respectively, by the factors $\exp(-2\pi\chi_r L/\lambda)$ and $\exp(-2\pi\chi_l L/\lambda)$, after the distance L. Defining by $\chi = (\chi_r + \chi_l)/2$ and $\chi' = (\chi_r - \chi_l)/2$ and substituting into (3.1) – (3.5) the real refraction indexes $n_l$ and $n_r$ by the complex ones $n_l$ and $n_r$ and following an analogous procedure used in the Sections (3.a)-(3.d) we get, instead of (3.7) the equation,

**E**(**r**,t)= $e_o$ Re{exp(π$\chi'$L/λ)(**x**+i**y**) exp(iδ) +

exp(–2π$\chi'$L/λ) (**x**–i**y**) exp(–iδ)} exp(iΦ´)    (4.1),

where  δ = π $(n_l - n_r)$L/λ   and   Φ´ = ω[t – $(n_l + n_r)$L/2c] + iπ$(\chi_r + \chi_l)$L/λ ].

From (4.1) we verify that the amplitude of the resulting wave, instead of (3.7) is now given by

2 $e_o$ {**x** cos(δ) – **y** sin(δ) }cosh(2π$\chi'$L/λ)–2 $e_o$ {**y** cos(δ)+**x** sin(δ) } sinh(2π$\chi'$L/λ) (4.2).

When $\chi_r = \chi_l$, that is, when $\chi' = 0$ we recover the wave amplitude given by (3.7):

2 $e_o${**x** cos(δ) – **y** sin(δ)}.
]

Indicating by **i**(δ)= **x** cos(δ) – **y** sin(δ) and **j**(δ)= – ( **x** sin(δ) + **y** cos(δ)) the unit vectors obtained rotating the unit vectors **i** and **j** by an angle δ > 0, that is,  in the clockwise sense, (4.2) is written as

2 $e_o$ {**i**(δ) cosh(2π$\chi'$L/λ) + **j**(δ) sinh(2π$\chi'$L/λ)} =  *a* **i**(δ) + *b* **j**(δ) ,     (4.3)

where we have put $a = 2\,e_o \cosh(2\pi\chi'L/\lambda)$ and $b = 2\,e_o \sinh(2\pi\chi'L/\lambda)$} that are the amplitudes of the wave along the vectors **i**(δ) and **j**(δ), respectively. In this way the resulting electric field, according to (4.1), will be the sum of two components, one along **i**(δ) and other along **j**(δ). They will oscillate in time with a frequency ω along these vectors with different amplitudes, *a* and *b*, respectively. In these conditions, as is known [8-16], the electric field describe an ellipse with axis *a* and *b*. The two components that initially entered forming a LP wave, after describing a distance L, now form an elliptically polarized wave. This phenomenon that is due to different absorptions of the RCP and LCP waves is named *circular dichroism* or *Cotton Effect*. As, in general, $\chi' = (\chi_r - \chi_l)/2$ is very small, about one part in one million [8-16], we have $a \gg b$, that is, the ellipse will be very flatten, becoming almost a straight segment with length *a*. So, the electric field will oscillate along a straight line that is rotated by an angle δ relatively to the vectors ***x*** and ***y***.

The ellipticity, which is equal to the ratio $b/a$ is conventionally measured by an angle Ψ using the following equation

$$\tan \Psi = b/a = \tanh(2\pi\chi'L/\lambda). \qquad (4.4).$$

Since $\chi'L \ll 1$, $\tan \Psi \approx \Psi$, (4.4) can be written as

$$\Psi \approx 2\pi\chi'L/\lambda = (\pi/\lambda)(\chi_r - \chi_l)L. \qquad (4.5).$$

In Section 6 we will calculate Ψ and show that the *rotatory power* and *circular dichroism* appear simultaneously.

### (5) Magnetic Birefringence (Faraday Effect).

Lets us show now using the classical atomic model that the action of a static magnetic field **B** applied on a dielectric medium is responsible for different refraction indexes of the RCP and LCP waves. This effect is called Faraday Effect or Magnetic Birefringence [2, 8 -11,13,15]. To see this in a simple way [10, 13] we assume that the dilute medium is composed by atoms with only one electron with mass m. It orbits the nucleus attracted by central force – K**r**, where $K = m\,\omega_o^2$ and $\omega_o$ is a constant angular velocity. We also assume that: (a) the incident CP wave has a frequency ω and propagates along the z-axis; (b) the circle described by the electron is in the (x,y) plane perpendicular to ***z*** and (c) the applied magnetic field is parallel to ***z*** , that is **B** = B ***z*** . In this geometric configuration the incident electric field **E** of CP wave, will be always in the plane of the circle described by the electron and given by **E** = E **r**/r. Under the action of the electric force – e**E** and of the central force – K**r** the electron will rotate with the angular velocity ω obeying the following equation:

$$- e\,\mathbf{E} - K\mathbf{r} = -m\,\omega^2\,\mathbf{r}.$$

It will be also assumed that there is no resonance between the incident wave and the atom, that is, $\omega \neq \omega_o$. The **E** angular velocity, has signal + or – if the wave is LCP or RCP, respectively. The orbital velocity of the electron would be given by **v** = ± ω r **θ** where **θ** is the unit vector tangent to the circle of radius r.

Since the static magnetic field **B** produces a Lorentz force $\mathbf{F}_B = -e\,\mathbf{v} \times \mathbf{B}$ on the electron we have the following equation of motion for the atomic electron,

$$- e\,\mathbf{E} \pm e\,B\,\omega\,\mathbf{r} - K\mathbf{r} = -m\,\omega^2\,\mathbf{r}.$$

Taking into account that $Kr = m\,\omega_o^2\,r$ we get from this equation,

$$\mathbf{r} = -(e/m)\,\mathbf{E}/(\omega_o^2 - \omega^2 \pm eB\omega/m),$$

where the signal + is valid for the LCP wave and the – for the *RCP* wave.

Indicating by N the number of atoms per unit volume and remembering that the atomic dipole moment $\mathbf{p} = -e\mathbf{r}$ the polarization vector **P** is given by

$$\mathbf{P} = (Ne^2/m)\,\mathbf{E}/(\omega_o^2 - \omega^2 \pm eB\omega/m). \qquad (5.1)$$

Note that the polarization **P** is directly proportional to the field **E** since we are assuming that dielectric medium is dilute. In the case of dense medium **P** would be proportional to an effective field $\mathbf{E}_{ef}$ according to (2.3).

Consequently,

$$\mathbf{D} = \mathbf{E} + 4\pi\,\mathbf{P} = \{1 + 4\pi(Ne^2/m)/(\omega_o^2 - \omega^2 \pm eB\omega/m)\}\,\mathbf{E}. \qquad (5.2)$$

Since $\mathbf{D} = \varepsilon\,\mathbf{E}$ we obtain,

$$\varepsilon = 1 + 4\pi\,(Ne^2/m)/(\omega_o^2 - \omega^2 \pm eB\omega/m). \qquad (5.3)$$

Taking into account that for usual dielectrics $n^2 = \varepsilon\mu \approx \varepsilon$ we verify that for the LCP and RCP the refraction indexes $n_l$ and $n_r$ are given, respectively, by:

$$\begin{aligned} n_l^2 &= 1 + 4\pi\,(Ne^2/m)/(\omega_o^2 - \omega^2 + eB\omega/m), \\ n_r^2 &= 1 + 4\pi\,(Ne^2/m)/(\omega_o^2 - \omega^2 - eB\omega/m), \end{aligned} \qquad (5.4)$$

From (5.4) we see that for **B** = 0 the refraction index n is given by,

$$n^2 = 1 + 4\pi\,(Ne^2/m)/(\omega_o^2 - \omega^2). \qquad (5.5)$$

Eqs. (5.4) – (5.5), that are valid for a dilute medium, show that $n^2$ is given by a relation $n^2 = 1 + 4\pi N \alpha$ where α is the polarizability defined by $\alpha_{\pm} = (e^2/m)/(\omega_o^2 - \omega^2 \pm eB\omega/m)$ and $\alpha = (e^2/m)/(\omega_o^2 - \omega^2)$.

Note that to deduce (5.4) we assumed that the light wave propagates in the same sense of **B**. In these conditions we obtained $n_r > n_l$ and consequently that rotatory power $[\alpha] = (\pi/\lambda)(n_l - n_r)$ defined by (3.8) is negative, that is, the medium is levorotatory or levogire. If the field **B** is inversed or inversing the propagation sense of the incident wave a contrary effect would be obtained, that is, the medium would become dextrorotatory or destrogire [8-10,13,15,16]. As the velocity of propagation of a wave is given by V = c/n, we see that the velocity $V_l = c/n_l > V_r = c/n_r$. Inversing **B** or the direction of propagation of the incident wave we would have the contrary, that is, $V_r > V_l$. Note that when **B** = 0 a substance naturally active, levogire or destrogire, does not change its optical activity if the direction of the wave is inversed.

So, we verified that due to the simultaneous action of the electric field **E** of the incident light wave and an applied static field **B** and two different refraction indexes, $n_r$ and $n_l$, are created. In next sections we will see that the simultaneous action of **E** and **B** of the incident light wave, depending on the molecular symmetries of the medium, can also produce $n_r \neq n_l$

**(6)Quantum Theory of the Molecular Optical Activity.**

Essentially, according to Biot and Pasteur [2,3], the refraction indexes $n_r$ and $n_l$ of the LCP and RCP waves would be different due to effects created by intrinsic properties of molecular symmetries of the medium where the light propagates. In this section we will see how to prove this conjecture. First, let us consider electromagnetic arguments to give an approximate idea of the effects produced by the incident field (**E,B**) on the molecules. After, using the Quantum Mechanics we will calculate rigorously these effects (**E,B**) determining [α] and $\Psi$.

From the classical electromagnetic point of view the incident field **E** polarizes the molecule: the positive charges move in the same sense of **E** and the negatives in the contrary sense, creating an *induced* electric dipole **p**. If the E is growing the charges move increasing their displacement inside the molecule tending to increase the dipole **p**. Suppose now that the molecular structure prevents the direct motion of the charges from the initial to a final position but that they are obliged, according to Pasteur, Biot and Drude (see details in the Lowry's book [2]) to move along helicoidal trajectories following however the **E** sense. This circulatory motion generates a magnetic field which is proportional to ∂**E**/∂t (Ampère's law) creating an *induced magnetic moment* **μ** ~ ∂**E**/∂t. The time dependent magnetic field **B** will produce a variable magnetic flux along the molecular helicoidal trajectories generating a charge displacement proportional to ∂**B**/∂t (Faraday's law). Since, according to Lenz's law, positive and negative charges move in contrary senses we must expect that the induced electric dipole will be given by **p** ≈ a **E** + b ∂**B**/∂t.

These arguments show that due to the molecular structures it becomes necessary to generalize the electromagnetic theory found in the traditional electromagnetic books like, for instance, of Panofsky & Phillips [5] and Jackson [6] to obtain correctly the electric polarization **P** and the magnetization **M** of the medium. It can be rigorously done only using the Quantum Mechanics as one can see in the works of Rosenfeld [17], Born [18,19] and Condon [12]. Following these authors we will calculate the induced electric and magnetic dipoles produced by a field (**E,B**) in a homogeneous dilute. Thus, let us consider a light wave (**E**, **B**) that propagates along the z-axis with **E** and **B** given by:

$$\mathbf{E}(\mathbf{r},t) = -(1/c)\, \partial \mathbf{A}(\mathbf{r},t)/\partial t \quad e \quad \mathbf{B}(\mathbf{r},t) = \text{rot } \mathbf{A}(\mathbf{r},t), \quad (6.1)$$

where

$$\mathbf{A}(\mathbf{r},t) = \text{Re}\{ \mathbf{A} \exp(i\Phi)\} = \{ \mathbf{A} \exp(i\Phi) + \mathbf{A}^* \exp(-i\Phi) \}/2$$

is the vector potential, $\Phi = \omega t - \mathbf{k}\cdot\mathbf{r} = \omega(t - n\mathbf{r}\cdot\mathbf{z}/c) = \omega(t - \mathbf{r}\cdot\boldsymbol{\sigma}/c)$, $\mathbf{k} = (2\pi/\lambda)\mathbf{z}$ is the wave vector, $\lambda = 2\pi c/\omega$ the wavelength in the vacuum and $\boldsymbol{\sigma} = n\mathbf{k}$.

Using (4.1) the fields **E** and **B** are written as,

$$\mathbf{E}(\mathbf{r},t) = -(i\omega/2c)\{ \mathbf{A} \exp(i\Phi) - \mathbf{A}^* \exp(-i\Phi)\}$$

e $\qquad\qquad\qquad\qquad\qquad\qquad\qquad\qquad\qquad\qquad\qquad\qquad (6.2)$

$$\mathbf{B}(\mathbf{r},t) = -(i\omega/2c)\{ (\boldsymbol{\sigma} \times \mathbf{A})\exp(i\Phi) - (\boldsymbol{\sigma} \times \mathbf{A}^*)\exp(-i\Phi)\}.$$

Neglecting the interaction of the light with the nucleus of the atoms the interaction potential $H$ of the electrons with the incident field is given by

$$H = -(e/mc) \sum_i \mathbf{A}(\mathbf{r}_i,t)\cdot\boldsymbol{\pi}_i = -(e/2mc) \sum_i \{\mathbf{A}(\mathbf{r}_i,t)\cdot\boldsymbol{\pi}_i + \boldsymbol{\pi}_i\cdot\mathbf{A}(\mathbf{r}_i,t)\} \quad (6.3),$$

where $\mathbf{r}_i$ and $\boldsymbol{\pi}_i = m\mathbf{v}_i$ are the positions and the linear momentum and velocities $\mathbf{v}_i$ of the i[th] electron. The interaction of the fields with the spins of the electrons, for simplicity, has not being considered. It will be done later.

Assuming that the wavelenghts $\lambda$ are much larger than the molecular dimensions r, that is, $\lambda \gg r$ we put $\exp(i\,\omega\boldsymbol{\sigma}\cdot\mathbf{r}/c) \approx 1 + i\omega\boldsymbol{\sigma}\cdot\mathbf{r}/c$. In these conditions the hamiltonian $H$ can be written as,

$$H = (1/2)[h\exp(i\omega t) + h^*\exp(-i\omega t)], \quad (6.4),$$

where

$$h = -(e/2mc)\sum_i \{ (\boldsymbol{\pi}_i\cdot\mathbf{A}) - (i\omega/2c)[(\boldsymbol{\pi}_i\cdot\mathbf{A})(\mathbf{r}_i\cdot\boldsymbol{\sigma}) + (\mathbf{r}_i\cdot\boldsymbol{\sigma})(\boldsymbol{\pi}_i\cdot\mathbf{A})]\}.$$

As the operators $(\boldsymbol{\pi}_i\cdot\mathbf{A})$ and $(\mathbf{r}_i\cdot\boldsymbol{\sigma})$ commute, that is, $(\boldsymbol{\pi}_i\cdot\mathbf{A})(\mathbf{r}\cdot\boldsymbol{\sigma}) - (\mathbf{r}_i\cdot\boldsymbol{\sigma})(\boldsymbol{\pi}_i\cdot\mathbf{A}) = 0$, the function h becomes

$$h = -(e/2mc)\sum_i \{ (\boldsymbol{\pi}_i\cdot\mathbf{A}) - (i\omega/c)[(\boldsymbol{\pi}_i\cdot\mathbf{A})(\mathbf{r}_i\cdot\boldsymbol{\sigma})]\}. \quad (6.5)$$

As $\boldsymbol{\pi}_i = m\mathbf{v}_i = m\, d\mathbf{r}_i/dt$, the electric **p** and magnetic **μ** dipole moments of the molecules are given, respectively, by

$$\mathbf{p} = \sum_i e\mathbf{r}_i \qquad \text{and} \qquad \boldsymbol{\mu} = \sum_i (e/2mc)(\mathbf{r}_i \times \boldsymbol{\pi}_i), \quad (6.6)$$

Using (6.6) we verify that the function h shown in (6.5) becomes,

$$h = -(1/2c)(d\mathbf{p}/dt)\mathbf{A} + (i\omega/c)\boldsymbol{\mu}\cdot(\boldsymbol{\sigma}\times\mathbf{A}). \quad (6.7)$$

To take into account the magnetization effect due to the electronic spins $S_i$ it is enough to substitute in (6.7) the magnetic moment **μ** by

$$\mu = \Sigma_i (e/2mc)\{ r_i \times \pi_i + 2S_i\}.$$

With the approximations seen in (6.5) – (6.7) the Schrödinger's equation that describes the interaction process of the light wave with the molecules of the medium is given by:

$$( i\hbar\, \partial\Psi/\partial t - H_o )\Psi = H\Psi , \qquad (6.8),$$

where $H_o$ is the hamiltonian of the free molecule and $H$ is the interaction hamiltonian described by (6.4)-(6.7).

Le us indicate by $\varphi_m$ and $E_m$, respectively, the eigenfunctions and eigenvalues of the operator $H_o$. Assuming that $H$ is a small perturbation of $H_o$ we can write the solution of (6.8), corresponding to the m.$^{th}$ stationary state $\Psi_m$ of the system in the form [20]:

$$\Psi_m = [\varphi_m + \Sigma_k (a_{km}\, \exp(i\omega t) + b_{km}\, \exp(-i\omega t)\, \varphi_k]\, \exp(-i\omega_m t) \qquad (6.9),$$

where $\omega_m = E_m/\hbar$.

Using (6.4),(6.7)-(6.9) and following the usual perturbation theory in quantum mechanics in the first order approximation [20] for the term h we calculate the wavefunctions $\Psi_m$ as seen in the Appendix. In these calculations we assume that the difference $\Psi_m - \varphi_m$ is small. In this way the obtained equations will be valid only when $\omega \neq \omega_{mk}$. Using these wavefunctions $\Psi_m$ the expected value $F_m$ of an observable **F** in the state $\Psi_m$ is given by [20]:

$$F_m = (-i/2\hbar) \Sigma_k \{ (\omega_{km}/\omega)^2 <m|F|k><k|p|m>\cdot \partial E/\partial t /(\omega_{mk}^2 - \omega^2)\}$$

$$+ (1/2\hbar) \Sigma_k \{ \omega_{km} <m|F|k><k|p|m>\cdot E/(\omega_{mk}^2 - \omega^2)\} +$$

$$+ (1/\hbar) \Sigma_k \{ \omega_{mk} <m|F|k><k|\mu|m>\cdot B/(\omega_{mk}^2 - \omega^2)\} +$$

$$+ (ic/\hbar) \Sigma_k \{ (\omega_{mk}/\omega)^2 <m|F|k><k|\mu|m>\cdot \partial B/\partial t /(\omega_{mk}^2 - \omega^2)\} \qquad (6.10).$$

In the case of electric dipole writing **F** = **p** and taking the real part of (6.10) we obtain $p_m = Re[F_m]$ :

$$p_m = (1/2\hbar) \Sigma_k \{(\omega_{km}/\omega)^2\, Im[<m|p|k><k|p|m>]\cdot \partial E/\partial t /(\omega_{mk}^2 - \omega^2)\}$$

$$+ (1/2\hbar)\Sigma_k \{\omega_{km}\, Re[<m|p|k><k|p|m>]\cdot E/(\omega_{mk}^2 - \omega^2)\}$$

$$+ (c/\hbar) \Sigma_k \{\omega_{mk}\, Re[<m|p|k><k|\mu|m>\}\cdot B/(\omega_{mk}^2 - \omega^2)\}$$

$$- (c/\hbar) \Sigma_k \{(\omega_{mk}/\omega)^2\, Im[<m|p|k><k|\mu|m>]\cdot \partial B/\partial t /(\omega_{mk}^2 - \omega^2)\} \qquad (6.11),$$

where we have taken into account that $<m|\mathbf{p}|k><k|\mathbf{p}|m> = |<m|\mathbf{p}|k>|^2$ is real and that $(\omega_{mk}/\omega)^2/(\omega_{mk}^2-\omega^2) = 1/\omega^2 + 1/(\omega_{mk}^2-\omega^2)$. So, we obtain

$\mathbf{p}_m = + (1/2\hbar)\Sigma_k \{\omega_{mk} \text{Re}[<m|\mathbf{p}|k><k|\mathbf{p}|m>]\cdot\mathbf{E}/(\omega_{mk}^2-\omega^2)\}$

$\quad + (c/\hbar) \Sigma_k \{\omega_{mk} \text{Re}[<m|\mathbf{p}|k><k|\boldsymbol{\mu}|m>\}\cdot\mathbf{B}/(\omega_{mk}^2-\omega^2)\}$

$\quad - (c/\hbar) \Sigma_k \{\text{Im}[<m|\mathbf{p}|k><k|\boldsymbol{\mu}|m>]\cdot \partial\mathbf{B}/\partial t /(\omega_{mk}^2-\omega^2)\}$ (6.12).

Analogously for the magnetic dipole, writing $\mathbf{F} = \boldsymbol{\mu}$ and neglecting terms of the order $\boldsymbol{\mu}^2$ in comparison with $|\mathbf{p}||\boldsymbol{\mu}|$ we get:

$\boldsymbol{\mu}_m = (-1/2\hbar) \Sigma_k \text{Im}\{<m|\boldsymbol{\mu}|k><k|\mathbf{p}|m>\partial\mathbf{E}/\partial t /(\omega_{mk}^2-\omega^2)\} +$

$\quad + (1/2\hbar) \Sigma_k \text{Re}\{\omega_{mk}<m|\boldsymbol{\mu}|k><k|\mathbf{p}|m>\cdot\mathbf{E}/(\omega_{mk}^2-\omega^2)\}$ (6.13).

The final values for the induced dipoles, electric and magnetic, are obtained doing an average over all possible orientations of the molecules in the space assuming that these directions are equally probable. To calculate the average, for instance, of the term involving the product [$\mathbf{p}\boldsymbol{\mu}\mathbf{B}$] over all orientations of $\mathbf{p}$ and $\boldsymbol{\mu}$ we calculate an average over all orientations of $\mathbf{p}$ and $\boldsymbol{\mu}$ taking constants the modulus of the dipoles and also fixed the angle between them. Choosing $\mathbf{B}$ as a reference axis to calculate all possible orientations of the vectors $\mathbf{p}$ and $\boldsymbol{\mu}$ we can show that [12] the average value [$\mathbf{p}\boldsymbol{\mu}\mathbf{B}$]$_{av}$ is given by dado $(1/3)[\mathbf{p}\boldsymbol{\mu}\mathbf{B}]$. Thus, the induced dipoles electric $\mathbf{d}_m$ and magnetic $\boldsymbol{\mu}_m$ become given by :

$$\mathbf{d_m} = \alpha_m \mathbf{E} + \gamma_m \mathbf{B} - (\beta_m/c) \partial\mathbf{B}/\partial t$$
$$\boldsymbol{\mu}_m = \gamma_m \mathbf{E} + (\beta_m/c) \partial\mathbf{E}/\partial t,$$
(6.14)

where the functions $\alpha_m$, $\beta_m$ and $\gamma_m$ for the states m are given by:

$\alpha_m = (2/3\hbar)\Sigma_k \{\omega_{km} |<m|\mathbf{p}|k>|^2 /(\omega_{mk}^2-\omega^2)\}$,

$\beta_m/c = (2/3\hbar) \Sigma_k \{\text{Im}[<m|\mathbf{p}|k>\cdot<k|\boldsymbol{\mu}|m>]/(\omega_{mk}^2-\omega^2)\}$, (6.15)

$\gamma_m = (2/3\hbar) \Sigma_k \{\omega_{mk} \text{Re}[<m|\mathbf{p}|k>\cdot<k|\boldsymbol{\mu}|m>]/(\omega_{mk}^2-\omega^2)\}$.

Using (6.14) and remembering that $\mathbf{D}$ is given by $\mathbf{D} = \mathbf{E} + 4\pi \mathbf{P}$ and that $\mathbf{B}$ is written as $\mathbf{B} = \mathbf{H} + 4\pi \mathbf{M}$, where $\mathbf{P} = N\mathbf{d}$ and $\mathbf{M} = N\boldsymbol{\mu}$, we have omitting, for simplicity, the index m:

$$\mathbf{D} = \varepsilon \mathbf{E} + \eta \mathbf{B} - (\rho/2c) \partial\mathbf{B}/\partial t$$
$$\mathbf{B} = \mathbf{H} + \eta \mathbf{E} + (\rho/2c) \partial\mathbf{E}/\partial t ,$$
(6.16)

where $\varepsilon = 1 + 4\pi N\alpha_m$, $\eta = 4\pi N\gamma_m$ and $\rho/2c = 4\pi N(\beta_m/c)$

Writing the fields **E**(**r**,t) and **B**(**r**,t) in the form **E**(**r**,t) = Re{**E**$_o$ exp(iΦ)} and **B** = Re{**B**$_o$ exp(iΦ)}, where Φ = ω(t − **σ**·**r** /c ) and **E**$_o$ e **B**$_o$ are constants, we obtain ∂**E**/∂t = iω**E** and ∂**B**/∂t = iω**B**. Using the rotational Maxwell equations rot(**E**) = (−1/c) ∂**B**/∂t e rot(**H**) = (1/c) ∂**D**/∂t, we get

$$-\boldsymbol{\sigma} \times \mathbf{H} = \mathbf{D} \quad \text{e} \quad \boldsymbol{\sigma} \times \mathbf{E} = \mathbf{B}. \qquad (6.17)$$

From (6.16) and (6.17) we get,

$$\mathbf{D} = -\boldsymbol{\sigma} \times \mathbf{H} = \varepsilon \mathbf{E} + \eta \mathbf{B} - (\rho/2c) \partial \mathbf{B}/\partial t$$
$$\mathbf{B} = \boldsymbol{\sigma} \times \mathbf{E} = \mathbf{H} + \eta \mathbf{E} + (\rho/2c) \partial \mathbf{E}/\partial t. \qquad (6.18)$$

From the second (6.18) relation we obtain **H** = **B** − η **E** − (ρ/2c) ∂**E**/∂t. Substituting this value in the first (6.18) relation and remembering that **σ** x **E** = **B**, we verify that

$$-\boldsymbol{\sigma} \times (\boldsymbol{\sigma} \times \mathbf{E}) = \varepsilon \mathbf{E} + (\rho/c)(\boldsymbol{\sigma} \times \partial \mathbf{E}/\partial t). \qquad (6.19).$$

As **σ** = n**z** = n k **k**, k = 2 π/λ and ∂**E**/∂t = i ω **E** from (6.19) we deduce

$$(n^2 - \varepsilon) \mathbf{E} = i (2\pi\rho/\lambda_n) (\mathbf{E} \times \mathbf{z}). \qquad (6.20).$$

Since **E** = E$_x$ **x** + E$_y$ **y** from (6.20) we obtain the following homogeneous system of equations

$$(n^2 - \varepsilon) E_x - i (2\pi\rho/\lambda_n) E_y = 0$$
$$i (2\pi\rho/\lambda_n) E_x + (n^2 - \varepsilon) E_y = 0. \qquad (6.29)$$

To avoid the trivial solution E$_x$ = E$_y$ = 0 the condition $(n^2 - \varepsilon)^2 = (2\pi\rho/\lambda_n)^2$ must be satisfied, that is, $n^2 = \varepsilon \pm 2\pi\rho/\lambda_n$ and E$_x$/ E$_y$ = ± i. Since ε >> 2πρ/λ$_n$ we have n ≈ √ε ± πρ/λ$_n$√ε. This implies that the RCP wave that has E$_x$/ E$_y$ = i would have a refraction index n$_r$ ≈ √ε + πρ/λ and the LCP wave which has E$_x$/ E$_y$ = − i would have a refraction index n$_l$ = √ε − πρ/λ.

According to the above calculations we see that n$_l$ − n$_r$ = 2πρ/λ. Since by (3.8) the rotatory power of the medium is given by [α] = (π/λ)(n$_l$ − n$_r$) we verify that

$$[\alpha] = (\pi/\lambda)(n_l - n_r) = (2\pi^2/\lambda^2) \rho. \qquad (6.30)$$

Since by (6.16) ρ = 8πcN(β$_m$/c) the rotatory power [α$_m$] for the m state according to (6.30) is written as

$$[\alpha_m] = (16\pi^3/\lambda^2) N_m \beta_m \qquad (6.31),$$

where N$_m$ = N P(m) is the average number of molecules that are in the φ$_m$ state and P(m) the corresponding Maxwell-Boltzmann distribution function.

Defining the "*rotational strength*" $R_{km}$ of the transition k → m by

$$R_{km} = \text{Im}\{<m|\mathbf{p}|k>\cdot<k|\mathbf{\mu}|m>\} \quad (6.32),$$

we see that the parameter $\beta_m$ given by (6.15) can be written as,

$$\beta_m = (2/3\hbar) \Sigma_k R_{km}/(\omega_{km}^2 - \omega^2)\}. \quad (6.33)$$

Finallly, using (6.31) the rotatory power $[\alpha_m]$ of the m state is given by

$$[\alpha_m] = (16\pi^3/\lambda_o^2) N_m (2/3\hbar) \Sigma_k R_{km}/(\omega_{km}^2 - \omega^2)\}. \quad (6.34)$$

As the effective β is given by the average value $\beta = \Sigma_m P(m) \beta_m$ the rotatory power [α] generated by the molecules of the dilute dielectric medium is written as,

$$[\alpha] = (16\pi^3/\lambda^2) N \beta. \quad (6.35)$$

We will show in next Section how the rotatory power [α] depends on the active molecular symmetries that are contained in the β parameter.

The (6.35) is valid for a dilute medium with only one kind of active molecule. When there are many kinds of different active molecules it is easy to see that for the mixture the following equation is valid:

$$[\alpha] = (16\pi^3/\lambda^2) \Sigma_r N_r \beta_r, \quad (6.36)$$

where $N_r$ and $\beta_r$ are, respectively, the density and the β parameter of the kind r molecule

In (6.35) it is assumed that the parameter β is a constant property of the individual molecule, practically independent of the medium where the molecule is immersed. If this occurs and the medium is dense (liquid or compressed gas) the rotatory power, instead of (6.35), is given by [12]:

$$[\alpha] = (16\pi^3/\lambda^2) N\beta(n^2 + 2)/3. \quad (6.37)$$

However, there are evidences [2] that β is very sensible to medium where the molecule is immersed, mainly with polar molecules and solvents. Note that, in this aspect, β is very different from the polarizability α which is practically independent of the medium.

**(7) Chiral Symmetry and Rotatory Power.**
According to (4.34) the rotatory power $[\alpha_m]$ is directly proportional to the factors $R_{km}$ defined by (4.32) that depend on the imaginary part of products of the electric and magnetic dipoles matrix elements. The $R_{km}$ have the following properties:

(A) $R_{km} = -R_{mk}$

(B) Let us indicate by P(x) the spatial inversion operator along a direction x so that P(x) f(x) = f(–x). If the molecule is symmetric along this direction we must have $P|m> = \pm|m>$ and $P|k> = \pm|k>$. Since P **d** = – **d** and P **μ** = **μ** the product **d μ** is a pseudoscalar and consequently $R_{km} = -R_{km} = 0$ giving [α] = 0. In this way we see that to have [α] ≠ 0 the molecule cannot have any plane of symmetry.

(C) Suppose now that $R_{km}$ ≠ 0 and that there are two molecules that are mirror images one of the other (chiral or dissymmetric molecules). As in this case $P|m> = |k>$ and $P|k> = |m>$ we see that $R_{Pk,Pm} = -R_{k,m}$. Consequently "chiral molecules have equal and opposed rotatory powers". In this way using (6.36), we verify that racemic solutions, that is, solutions with 50% of L molecules and with 50% of R molecules have [α] = 0.

Other interesting property of the rotatory power is obtained taking into account that $\Sigma_k R_{k,m} = 0$, that is,

$$\Sigma_k R_{k,m} = \text{Im} \{\Sigma_k <m|\mathbf{p}|k> \cdot <k|\mathbf{\mu}|m>\} = \text{Im} \{<m|\mathbf{p}\cdot\mathbf{\mu}|m>\} = 0,$$

because the diagonal matrix of any observable is always real [20]. Thus, for very high frequencies, that is, $\omega \gg \omega_{k,m}$ the parameter $\beta_m$ defined by (4.35) becomes

$$\beta_m \approx -(2/3\hbar\omega^2) \; \Sigma_k R_{km} = 0.$$

In other words, for very high frequencies or very short wavelengths ($\lambda_{km} \ll \lambda$) the rotatory power tends to zero. On the other side, as [α] = $(16\pi^3 N \beta)/\lambda^2$, we verify that [α] also tends to zero for very large or very short wavelengths λ. That is, [α] tends to zero in the both extremes of the spectrum, that is, for very short and for very long waves.

**(8) Equations for α, β and γ for resonance case $\omega = \omega_{km}$.**

Eqs.(6.15) for the parameters α, β and γ are valid only for frequencies $\omega \neq \omega_{km} = \omega_k - \omega_m$. They cannot be used for resonant cases, that is, when $\omega \approx \omega_{km}$. To treat resonances it is necessary to take into account the average lifetime Γ of the excited states. In this case [20] the amplitude $a_m(t)$ of a state $\varphi_m$ is represented by $a_m(t) = \exp(-i E_m t /\hbar) \exp(-\Gamma_m t /2)$. Formally we take into account the average lifetime $\Gamma_m$ of the excited state $\varphi_m$ transforming the energy $E_m$ in a complex number, $E_m \rightarrow E_m - i\hbar\Gamma_m/2$. Performing a perturbative calculation in the first order approximation similar to that shown in Section (6) we obtain [20]

$$\alpha_m \approx (2/3\hbar) \{ \Sigma_{k \neq m} |<m|\mathbf{p}|k>|^2 \, \omega_{mk} / [\omega_{mk}^2 - \omega^2 + i\omega\Gamma_{km}/2] \},$$

(8.1)

$$\beta_m \approx (2c/3\hbar) \{\Sigma_{k \neq m} R_{km}/ [\omega_{mk}^2 - \omega^2 + i\omega\Gamma_{km}/2] \},$$

$$\gamma_m \approx (2/3\hbar) \Sigma_k \{\omega_{mk} \text{Re}[<m|\mathbf{p}|k> \cdot <k|\mathbf{\mu}|m>] / [\omega_{mk}^2 - \omega^2 + i\omega\Gamma_{km}/2] \}.$$

where $\Gamma_{km} = \Gamma_k + \Gamma_m$. To deduce (8.1), taking into account that $\Gamma$ values are very small ($\Gamma \sim 10^{-8}$ s for optical and $\Gamma \sim 10^{-7}$ s for microwave transitions) compared with other terms that appear in the numerators, we neglected $\Gamma$ in the numerators. In the denominators we have considered only the first order $\Gamma$ terms neglecting the second order $\Gamma^2$ terms.

Note that in the limit $\Gamma \to 0$ the parameters $\alpha_m$, $\beta_m$ and $\gamma_m$ shown in (8.1) become real given by (6.15).

The complex dielectric constant $\varepsilon$ can be obtained [20] substituting $\alpha$ defined by (8.1) into (2.7) giving $\varepsilon \approx 1 + 4\pi N \alpha$. With the analytic extension $\varepsilon \approx n^2$ where $n = n + i \chi$ we can get the real (n) and imaginary ($\chi$) parts of $n$ with the following equation [20]:

$$\varepsilon \approx n^2 = (n + i\chi)^2 = 1 + 4\pi N \alpha. \qquad (8.2)$$

As we are interested in the rotatory power and circular dichroism we will analyze only relations that depend of the parameter $\beta_m$. So, decomposing $\beta_m$ defined by (8.1) into its real and imaginary parts,

$$\text{Re}\{\beta_m\} = (2c/3\hbar) \Sigma_{k \neq m} \{ R_{km} (\omega_{mk}^2 - \omega^2) / [(\omega_{mk}^2 - \omega^2)^2 + \omega^2 \Gamma_{km}^2/4] \},$$

$$\text{Im}\{\beta_m\} = (c/3\hbar) \Sigma_{k \neq m} \{ R_{km} \omega \Gamma_{km} / [(\omega_{mk}^2 - \omega^2)^2 + \omega^2 \Gamma_{km}^2/4] \}, \qquad (8.3)$$

we define the *complex rotatory power* $[\Theta_m]$ :

$$[\Theta_m] = (16\pi^3/\lambda^2) N \{\text{Re}[\beta_m] + i \text{Im}[\beta_m]\}. \qquad (8.4)$$

In these conditions the rotatory power $[\alpha_m]$, instead (6.31) is now given by

$$[\alpha_m] = (16\pi^3/\lambda^2) N \text{Re}[\beta_m] . \qquad (8.5)$$

Taking into account the complex refraction index $n_l = n_l + i\chi_l$ and $n_r = n_r + i\chi_r$, it is easy to see that with a complex representation the *complex rotatory power* $[\Theta]$ is given by (omitting the indices m to simplify the notation):

$$[\Theta] = (\pi/\lambda)(n_l - n_r) = (\pi/\lambda)\{(n_l - n_r) + i(\chi_l - \chi_r)\} = [\alpha] + i[\Psi], \qquad (8.6)$$

where $[\alpha] = (\pi/\lambda)(n_l - n_r)$ and $[\Psi] = (\pi/\lambda)(\chi_l - \chi_r)$. The first term of this complex function $[\Theta]$ gives the rotation angle of the polarization plane per unit of length $[\alpha] = \delta/L$, according to (3.8). The second term $[\Psi]$, remembering that the ellipticity $\Psi \approx (\pi/\lambda)(\chi_l - \chi_r)L$ (see (4.5)), is given by $[\Psi] = \Psi/L$ which measures the ellipticity $\Psi$ per unit of length.

Many experimental results compared with the theoretical predictions given by (8.3)-(8.6) are seen in several books and works [2,3,9-13,15,16]. There is an excellent agreement between theory and experiment.

According to (8.3)-(8.6) the rotatory power $[\Theta]$ is a complex function of the frequency $\omega$, that is, $[\Theta] = [\Theta(\omega)] = [\alpha(\omega)] + i[\Psi(\omega)]$. In this way, using the analytic properties of the complex function $[\Theta(\omega)]$ it can be shown the following relations connecting its real and imaginary parts [6]:

$$[\alpha(\omega)] = 1 + (2/\pi) P \int_o^\infty d\omega'\, \omega'\, [\varphi'(\omega)]/(\omega'^2 - \omega^2)$$

$$[\Psi(\omega)] = -(2\omega/\pi) P \int_o^\infty d\omega'\, ([\varphi(\omega)] - 1)/(\omega'^2 - \omega^2),$$

(8.7)

where the symbol P indicates the *principal value* of the integral. The (8.7) that are known as *Kramers-Krönig* dispersion relations show that mathematically that the rotatory power and the Condon Effect are intrinsically connected: one effect does not exist without the other.

**(9) Drude Equation and Biot Law.**

Let us first analyze the simplest case when the active molecules have only one resonance frequency $\omega_{mk} = \omega_o$ in the neighborhood of the incident frequency $\omega$. To do this we will adopt a procedure commonly used in many optics books that is to write $[\Theta]$ as a function of the wavelengths $\lambda = 2\pi c/\omega$ instead of the frequencies $\omega$. So, using (8.3)-(8.5) the complex rotatory power $[\Theta]$ is written as [9] :

$$[\Theta(\lambda)] = [\alpha(\lambda)] + i[\Psi(\lambda)], \quad (8.7)$$

where 
$$[\alpha(\lambda)] = A(\lambda^2 - \lambda_o^2)/[(\lambda^2 - \lambda_o^2)^2 + \lambda^2 \lambda_o^2 g^2] \quad \text{and}$$

$$[\Psi(\lambda)] = A\Gamma\lambda / [(\lambda^2 - \lambda_o^2)^2 + \lambda^2 \lambda_o^2 g^2]$$

where $A = (8\pi/3c\hbar)NR_o\lambda_o^2$, $R_o = \text{Im}\{<m|\mathbf{p}|k>\cdot<k|\mathbf{\mu}|m>\}$ that takes into account the transition $\varphi_m \to \varphi_k$ between two defined states with $\omega_{mk} = \omega_o$ and
$g = 2\pi\lambda_o \Gamma_o/c$. The parameter $\Gamma_o = \Gamma_k + \Gamma_m$ gives the width of the line, measured in rad/s, in the transition $\varphi_m \to \varphi_k$. Note that in (8.7) g is an adimensional factor.

In Fig.4 is shown the generic behavior of $[\alpha(\lambda)]$ and $[\Psi(\lambda)]$ as a function of the wavelength $\lambda$ assuming A positive. When A is positive $[\Psi(\lambda)]$ is always positive and the Cotton Effect is called *positive* Cotton Effect. When A is negative $[\Psi(\lambda)]$ is always negative and we have the *negative* Cotton Effect.

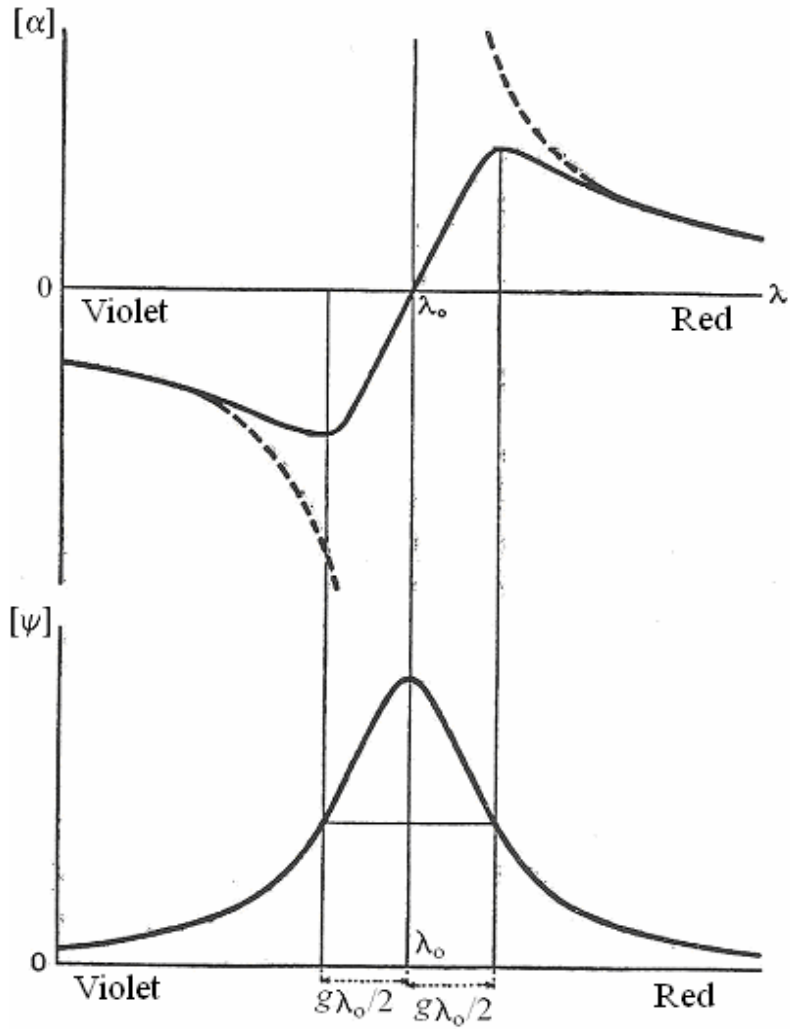

Figure 4. The rotatory power [α(λ)] and the ellipticity parameter [Ψ(λ)] as a function of the wavelength λ.

Far from the resonance, that is, when $\lambda \neq \lambda_{ok}$, the rotatory power [α(λ)] can be written as

$$[\alpha(\lambda)] = \Sigma_k \ A_k /(\lambda^2 - \lambda_{ok}^2), \qquad (8.8)$$

which is a relation known as Drude Equation. For very large incident wavelengths $\lambda \gg \lambda_{ok}$ the (8.8) assumes a very simple form

$$[\alpha(\lambda)] = A/\lambda^2 , \qquad (8.9)$$

known as Biot Law.

**Appendix.**

As was seen at the beginning of Sec.6 the Schrödinger equation that describes the interaction process of the incident light wave (**E,B**) with the molecules of the medium is given by

$$( i\hbar\, \partial\Psi/\partial t - H_o )\Psi = H\Psi , \qquad (A.1)$$

where $H_o$ is the hamiltonian of the non perturbed molecule and $H$ is the interaction hamiltonian given by

$$H = (1/2)[h\, \exp(i\omega t) + h^*\, \exp(-i\omega t)] \qquad (A.2),$$

where

$$h = - (1/2c)\, (d\mathbf{p}/dt) \cdot \mathbf{A} + (i\omega/c)\, \boldsymbol{\mu} \cdot (\boldsymbol{\sigma} \times \mathbf{A}), \qquad (A.3)$$

with $\mathbf{p} = \Sigma_i e\mathbf{r}_i$, $\boldsymbol{\mu} = \Sigma_i (e/2mc)\{\mathbf{r}_i \times \boldsymbol{\pi}_i + \mathbf{S}_i\}$ e $\boldsymbol{\pi}_i = m\, \mathbf{v}_i = m\, d\mathbf{r}_i/dt$.

Let us indicate by $\varphi_m$ and $E_m$, respectively, the eigenfunctions and eigenstates as of the operator $H_o$. Considering $H$ as a small perturbation of $H_o$ we can write [20] the solutions of (6.8) for the m[th] stationary state $\Psi_m$ of the system as

$$\Psi_m = [\varphi_m + \Sigma_k (a_{km} \exp(i\omega t) + b_{km} \exp(-i\omega t)\, \varphi_k ]\, \exp(-i\omega_m t), \qquad (A.4)$$

where $\omega_m = E_m/\hbar$.

Substituting (A.4) into (A.1) and using (A.2) we obtain [20], in a first order approximation, the following equation:

$$\Sigma_k [\, (\omega_{mk} - \omega)\, a_{km} \exp(i\omega t) + (\omega_{mk} + \omega)\, a_{km} \exp(-i\omega t)]\, \varphi_k =$$

$$= (1/2\hbar)(h \exp(i\omega t) + h^* \exp(-i\omega t))\, \varphi_m , \qquad (A.5)$$

with $\omega_{mk} = \omega_m - \omega_k$.

Multiplying both sides of (A.5) by $\varphi_k^*$ and integrating over all values of the arguments of the eigenfunctions $\varphi$ we obtain (making equal the coefficients of the exponential factors with the same temporal arguments) the following values for the coefficients to be determined $a_{km}$ and $b_{km}$ of (A.5):

$$a_{km} = <k|h|m>/[2\hbar(\omega_{mk} - \omega)] \quad \text{and} \quad b_{km} = <k|h^*|m>/[2\hbar(\omega_{mk} + \omega)], \qquad (A.6)$$

where $<k|f|m> = \int \varphi_k^*\, f\, \varphi_m\, d^3\mathbf{r}$ with $f = h$ or $h^*$.

In the above calculations we assumed that the difference $\Psi_m - \varphi_m$ is very small which implies that the (A.6) can be used only when $\omega \neq \omega_{mk}$

From (A.4) and (A.6) we verify that the expected value $\mathbf{F}_m = <\Psi_m|\mathbf{F}|\Psi_m>$ of an observable **F** for the state $\Psi_m$ is given by [20]

$$<\Psi_m|\mathbf{F}|\Psi_m> = <\varphi_m|\mathbf{F}|\varphi_m> + \{<m|\mathbf{F}|k><k|h|m> \exp(i\omega t)\, /[2\hbar(\omega_{mk} - \omega)]$$

$$+ <m|\mathbf{F}|k><k|h^*|m> \exp(-i\omega t) /[2\hbar(\omega_{mk} + \omega)] \}. \quad (A.7)$$

The first term $<\varphi_m|\mathbf{F}|\varphi_m>$ gives the expected value **F** for the non perturbed state $\varphi_m$ and the second one that depends on h and h* describes the perturbation created by the incident wave on the molecule. Since $\mathbf{F} = \mathbf{p}$ or $\boldsymbol{\mu}$ the first term gives the permanent dipoles (electric or magnetic) for the state $\varphi_m$ and the second gives the induced dipoles that are calculated using (4.12) taking into account the effects of all states $\varphi_k$. Assuming that the states $\varphi_m$ have defined parities the permanent electric dipoles are equal to zero since $<\varphi_m|\mathbf{p}|\varphi_m> = 0$. It will be also assumed that the dielectric medium has negligible magnetic properties composed by molecules with null permanent magnetic dipoles, that is, $<\varphi_m|\boldsymbol{\mu}|\varphi_m> = 0$. In these conditions we have only the contributions of the induced dipoles $\mathbf{p}_m$ and $\boldsymbol{\mu}_m$ that are calculated using the above equation:

$$\mathbf{F}_m = \Sigma_k \{<m|\mathbf{F}|k><k|h|m> \exp(i\omega t) /[2\hbar(\omega_{mk} - \omega]$$

$$+ <m|\mathbf{F}|k><k|h^*|m> \exp(-i\omega t) /[2\hbar(\omega_{mk} + \omega)] \}. \quad (A.8)$$

Taking into account that $<k|d\mathbf{p}/dt|m> = i\omega_{km}<k|\mathbf{p}|m>$ we see that $<m|h|k>$ and $<m|h^*|k>$, using (A.3), are given by

$$<k|h|m> = -(i/2c)\omega_{km}<k|\mathbf{p}|m>\cdot\mathbf{A} + (i\omega/2c)<k|\boldsymbol{\mu}|m>\cdot(\boldsymbol{\sigma} \times \mathbf{A})$$

and $\hspace{10cm}$ (A.9)

$$<k|h^*|m> = -(i/2c)\omega_{km}<k|\mathbf{p}|m>\cdot\mathbf{A} - (i\omega/2c)<k|\boldsymbol{\mu}|m>\cdot(\boldsymbol{\sigma} \times \mathbf{A}).$$

Substituting (A.9) into (A.8) results:

$$\mathbf{F}_m = (-i/2c)\exp(i\omega t)\Sigma_k\{<m|\mathbf{F}|k><k|\mathbf{p}|m>\cdot\mathbf{A}\,\omega_{km}/[2\hbar(\omega_{mk} - \omega)]\}$$

$$+ (i\omega/2c)\exp(i\omega t)\Sigma_k\{<m|\mathbf{F}|k><k|\boldsymbol{\mu}|m>\cdot(\boldsymbol{\sigma} \times \mathbf{A})]/[2\hbar(\omega_{mk} - \omega)]$$

$$- (i/2c)\exp(-i\omega t)\Sigma_k\{<m|\mathbf{F}|k><k|\mathbf{p}|m>\cdot\mathbf{A}\,\omega_{km}/[2\hbar(\omega_{mk} + \omega)]\}$$

$$- (i\omega/2c)\exp(-i\omega t)\Sigma_k\{<m|\mathbf{F}|k><k|\boldsymbol{\mu}|m>\cdot(\boldsymbol{\sigma} \times \mathbf{A})]./[2\hbar(\omega_{mk} + \omega)]\}$$

$$= (-i/2c)\Sigma_k\{<m|\mathbf{F}|k><k|\mathbf{p}|m>\cdot\mathbf{A}[\omega_{km}\exp(i\omega t)/2\hbar(\omega_{mk} - \omega] -$$

$$\exp(-i\omega t)\omega_{km}/[2\hbar(\omega_{mk} + \omega)]\}$$

$$+(i\omega/2c)\Sigma_k\{<m|\mathbf{F}|k><k|\boldsymbol{\mu}|m>\cdot(\boldsymbol{\sigma} \times \mathbf{A})](\exp(i\omega t)/[2\hbar(\omega_{mk} - \omega)]$$

$$-\Sigma_k\{<m|\mathbf{F}|k><k|\boldsymbol{\mu}|m>\cdot(\boldsymbol{\sigma} \times \mathbf{A})[(\exp(-i\omega t)/[2\hbar(\omega_{mk} + \omega)]\}.$$

We can also write $\mathbf{F}_m$ in the form

$\mathbf{F}_m = (-i/4c\hbar) \Sigma_k \{ \omega_{km} <m|\mathbf{F}|k><k|\mathbf{p}|m> \cdot$

$\qquad \mathbf{A} [ \exp(i\omega t)/(\omega_{mk} - \omega) - \exp(-i\omega t)/(\omega_{mk} + \omega)] \} +$

$(i\omega/2c\hbar) \Sigma_k \{<m|\mathbf{F}|k><k|\mathbf{\mu}|m> \cdot$

$\qquad (\mathbf{\sigma} \times \mathbf{A}) [ \exp(i\omega t)/(\omega_{mk} - \omega) - \exp(-i\omega t)/(\omega_{mk} + \omega)] \}$  (A.10).

Since $\exp(i\omega t)/(\omega_{mk} - \omega) - \exp(-i\omega t)/(\omega_{mk} + \omega) =$

$\qquad = [(\omega_{mk} + \omega) \exp(i\omega t) - (\omega_{mk} - \omega) \exp(-i\omega t)]/(\omega_{mk}^2 - \omega^2) =$

$\qquad = \{\omega_{mk}(\exp(i\omega t) - \exp(-i\omega t)) + \omega(\exp(i\omega t) + \exp(-i\omega t))\}/(\omega_{mk}^2 - \omega^2)$

$\qquad = \{2 i\omega_{mk} \sin(\omega t) + 2 \omega \cos(\omega t)\}/(\omega_{mk}^2 - \omega^2)$

(A.10) becomes

$\mathbf{F}_m = (-i/4c\hbar) \Sigma_k \{ \omega_{km} <m|\mathbf{F}|k><k|\mathbf{p}|m> \cdot$

$\qquad \mathbf{A} [2 i\omega_{mk} \sin(\omega t) + 2 \omega \cos(\omega t)]/(\omega_{mk}^2 - \omega^2)\}$

$+ (i\omega/2c\hbar) \Sigma_k \{<m|\mathbf{F}|k><k|\mathbf{\mu}|m> \cdot$

$\qquad (\mathbf{\sigma} \times \mathbf{A}) [2\omega \cos(\omega t) + 2 i \omega_{mk} \sin(\omega t)]/(\omega_{mk}^2 - \omega^2)\}$  (A.11).

Taking into account the dipolar approximation that was used to obtain the function h, the vector potential $\mathbf{A}$ given by (A.3) can be written $\mathbf{A} = \text{Re}\{\mathbf{A}\exp(i\Psi)\} \approx \text{Re}\{\mathbf{A}\exp(i\omega t) + \mathbf{A}^*\exp(-i\omega)\}/2 = \mathbf{A}\cos(\omega t)$, where $\mathbf{A}$ is real. Consequently, $\mathbf{E}(t) = (-1/c) \partial\mathbf{A}(t)/\partial t = (\omega/c) \mathbf{A} \sin(\omega t)$, $\mathbf{B}(t) = -(\omega/c)(\mathbf{\sigma} \times \mathbf{A})\sin(\omega t)$, $\partial\mathbf{E}(t)/\partial t = (\omega^2/c) \mathbf{A} \cos(\omega t)$ and $\partial\mathbf{B}(t)/\partial t = (\omega^2/c)(\mathbf{\sigma} \times \mathbf{A})\cos(\omega t)$. Using these results we get

$\mathbf{F}_m = (-i/2c\hbar) \Sigma_k \{ \omega_{km} <m|\mathbf{F}|k><k|\mathbf{p}|m> \cdot$

$\qquad [(c\omega_{km}/\omega^2) \partial\mathbf{E}(t)/\partial t + (ic/\omega) \mathbf{E}(t)]/(\omega_{mk}^2 - \omega^2)\} \cdot +$

$+ (i\omega/c\hbar) \Sigma_k \{<m|\mathbf{F}|k><k|\mathbf{\mu}|m> \cdot$

$\qquad [(c/\omega) \partial\mathbf{B}(t)/\partial t - (ic\omega_{mk}/\omega) \mathbf{B}(t)]/(\omega_{mk}^2 - \omega^2)\}$  (A.12).

That is,

$\mathbf{F}_m = (-i/2\hbar) \Sigma_k \{ (\omega_{km}/\omega)^2 <m|\mathbf{F}|k><k|\mathbf{p}|m> \cdot \partial\mathbf{E}/\partial t /(\omega_{mk}^2 - \omega^2)\}$

$+ (1/2\hbar) \Sigma_k \{ \omega_{km} <m|\mathbf{F}|k><k|\mathbf{p}|m> \cdot \mathbf{E}/(\omega_{mk}^2 - \omega^2)\} +$

$$+ (1/\hbar) \Sigma_k \{ \omega_{mk} <m|\mathbf{F}|k><k|\mathbf{\mu}|m>\cdot \mathbf{B}/ (\omega_{mk}^2- \omega^2)\} +$$

$$+ (ic/\hbar) \Sigma_k \{ (\omega_{mk}/\omega)^2 <m|\mathbf{F}|k><k|\mathbf{\mu}|m>\cdot \partial \mathbf{B}/\partial t / (\omega_{mk}^2- \omega^2)\}. \quad (A.13)$$

For the electric dipole case, $\mathbf{F} = \mathbf{p}$, we obtain taking the real part of (A.13), that is, making $\mathbf{p}_m = Re[\mathbf{F}_m]$:

$$\mathbf{p}_m = (1/2\hbar) \Sigma_k \{(\omega_{km}/\omega)^2 \, Im[<m|\mathbf{p}|k><k|\mathbf{p}|m>]\cdot \partial \mathbf{E}/\partial t /(\omega_{mk}^2- \omega^2)\}$$

$$+ (1/2\hbar)\Sigma_k \{\omega_{km} \, Re[<m|\mathbf{p}|k><k|\mathbf{p}|m>]\cdot \mathbf{E}/ (\omega_{mk}^2- \omega^2)\}$$

$$+ (c/\hbar) \Sigma_k \{\omega_{mk} \, Re[<m|\mathbf{p}|k><k|\mathbf{\mu}|m>\}\cdot \mathbf{B}/ (\omega_{mk}^2- \omega^2)\}$$

$$- (c/\hbar) \Sigma_k \{(\omega_{mk}/\omega)^2 \, Im[<m|\mathbf{p}|k><k|\mathbf{\mu}|m>]\cdot \partial \mathbf{B}/\partial t / (\omega_{mk}^2- \omega^2)\} \quad (A.14),$$

where we used the fact that $<m|\mathbf{p}|k><k|\mathbf{p}|m> = |<m|\mathbf{p}|k>|^2$ is real and that $(\omega_{mk}/\omega)^2/(\omega_{mk}^2- \omega^2) = 1/\omega^2 + 1/(\omega_{mk}^2- \omega^2)$. So, we obtain

$$\mathbf{p}_m = + (1/2\hbar)\Sigma_k \{\omega_{mk} \, Re[<m|\mathbf{p}|k><k|\mathbf{p}|m>]\cdot \mathbf{E}/ (\omega_{mk}^2- \omega^2)\}$$

$$+ (c/\hbar) \Sigma_k \{\omega_{mk} \, Re[<m|\mathbf{p}|k><k|\mathbf{\mu}|m>\}\cdot \mathbf{B}/ (\omega_{mk}^2- \omega^2)\}$$

$$- (c/\hbar) \Sigma_k \{ Im[<m|\mathbf{p}|k><k|\mathbf{\mu}|m>]\cdot \partial \mathbf{B}/\partial t / (\omega_{mk}^2- \omega^2)\} \quad (A.15).$$

Similarly, for the magnetic dipole, $\mathbf{F} = \mathbf{\mu}$, neglecting second order terms $\mathbf{\mu}^2$ relatively to $|\mathbf{p}||\mathbf{\mu}|$, we get:

$$\mathbf{\mu}_m= (-1/2\hbar) \Sigma_k \, Im\{<m|\mathbf{\mu}|k><k|\mathbf{p}|m>\cdot \partial \mathbf{E}/\partial t /(\omega_{mk}^2- \omega^2)\} +$$

$$+ (1/2\hbar) \Sigma_k \, Re\{ \omega_{mk}<m|\mathbf{\mu}|k><k|\mathbf{p}|m>\cdot \mathbf{E}/ (\omega_{mk}^2- \omega^2)\} . \quad (A.16)$$

The final result for the induced dipoles, electric and magnetic, is obtained doing an average over all possible orientations of the molecules in the space assuming that all directions are equally probable. To calculate the average, for instance, of the product [**pμB**] over all orientations of **p** and **μ** we fix the modulus of the dipoles and also the angle between them. Choosing **B** as a reference axis to calculate all possible orientations of the vectors **p** and **μ** we can show [12] that the average value [**pμB**]$_{av}$ is given by (1/3)[**pμB**].

In these conditions the induced electric $\mathbf{d}_m$ and magnetic $\mathbf{m}_m$ dipoles are given by,

$$\mathbf{d_m} = \alpha_m \, \mathbf{E} + \gamma_m \, \mathbf{B} - (\beta_m/c) \, \partial \mathbf{B}/\partial t$$

$$\mathbf{m}_m = \gamma_m \, \mathbf{E} + (\beta_m/c) \, \partial \mathbf{E}/\partial t, \quad (A.17)$$

where the functions $\alpha_m$, $\beta_m$ and $\gamma_m$ for the states m are given by

$$\alpha_m = (2/3\hbar)\Sigma_k \{ \omega_{km} |<m|\mathbf{p}|k>|^2 /(\omega_{mk}^2 - \omega^2) \},$$

$$\beta_m /c = (2/3\hbar) \Sigma_k \{ \text{Im}[<m|\mathbf{p}|k> \cdot <k|\mathbf{\mu}|m>] /(\omega_{mk}^2 - \omega^2) \}, \qquad (A.18)$$

$$\gamma_m = (2/3\hbar) \Sigma_k \{\omega_{mk} \text{Re}[<m|\mathbf{p}|k> \cdot <k|\mathbf{\mu}|m>] / (\omega_{mk}^2 - \omega^2) \}.$$